\documentclass[letterpaper, 10 pt, conference]{ieeeconf}  

\IEEEoverridecommandlockouts                              

\usepackage{graphics} 
\usepackage{cite}
\usepackage{amsmath,amssymb,amsfonts}
\usepackage{pifont}
\usepackage{algorithmic}
\usepackage{graphicx}
\usepackage{textcomp}
\usepackage{xcolor}
\usepackage{booktabs}
\usepackage{multirow}
\usepackage{makecell}
\usepackage{tabularx}
\usepackage{hyperref}
\usepackage{orcidlink}
\usepackage{subfigure}

\title{\LARGE \bf
Contrastive Learning-based User Identification \\ with Limited Data on Smart Textiles
}

\author{Yunkang Zhang$^{1}$, Ziyu Wu$^{2}$, Zhen Liang$^{3}$, Fangting Xie$^{4}$, Quan Wan$^{5}$, Mingjie Zhao$^{6}$ and Xiaohui Cai$^{*}$
\thanks{This work is supported by the National Natural Science Foundation of
China under Grant No. 62072420. }
\thanks{*Corresponding author. All authors are with the University of Science and Technology of China, Huangshan Road 443, Dian 3 buliding, Hefei, 230027, China~(email: ykzhang, wzy1999, LZ1999, xieft, wanquan2020, zmj4527@mail.ustc.edu.cn, caixiaohui@ustc.edu.cn)}
}

\begin{document}

\maketitle
\thispagestyle{empty}
\pagestyle{empty}

\begin{abstract}

Pressure-sensitive smart textiles are widely applied in the fields of healthcare, sports monitoring, and intelligent homes. The integration of devices embedded with pressure sensing arrays is expected to enable comprehensive scene coverage and multi-device integration. However, the implementation of identity recognition, a fundamental function in this context, relies on extensive device-specific datasets due to variations in pressure distribution across different devices. To address this challenge, we propose a novel user identification method based on contrastive learning. We design two parallel branches to facilitate user identification on both new and existing devices respectively, employing supervised contrastive learning in the feature space to promote domain unification. When encountering new devices, extensive data collection efforts are not required; instead, user identification can be achieved using limited data consisting of only a few simple postures. Through experimentation with two 8-subject pressure datasets (BedPressure and ChrPressure), our proposed method demonstrates the capability to achieve user identification across 12 sitting scenarios using only a dataset containing 2 postures. Our average recognition accuracy reaches 79.05\%, representing an improvement of 2.62\% over the best baseline model.

\end{abstract}


\section{INTRODUCTION}

In recent years, the development of pressure-sensitive smart textiles has brought new vitality to fields such as healthcare~\cite{id3}, sports monitoring~\cite{b1}, and beyond~\cite{luo1, id6, luo2}. It is foreseeable that in the future, a wide range of devices will be integrated with smart textiles to provide diverse services. As an essential function of smart textiles, user identification holds significant potential for practical applications. For example, Muhammad et al.~\cite{b6} proposed the use of smart textile shirts for biometric authentication through electrocardiogram recognition, enabling personalized heart monitoring. Zhang et al.~\cite{b7} utilized smart sensing seats for user identification, offering personalized sitting detection and allowing adjustments to parameters such as temperature, height, and softness.

However, unlike fingerprint and face recognition technologies, which have standardized datasets and algorithms available across different devices and scenarios, pressure-sensitive devices produce richer data with significant variations in pressure distribution across different devices and postures. As a result, each new device requires extensive data collection covering diverse scenarios for effective user identification implementation. To mitigate the reliance on large-scale datasets, it is desirable to develop a universal method for various pressure-sensitive smart textiles. Nevertheless, due to existing domain gaps, developing universal identity recognition algorithms for pressure-sensitive smart textiles remains challenging.

To the best of our knowledge, there is currently no existing work that achieves cross-device identification for smart textiles. Therefore, we adopt an alternative approach to reduce reliance on extensive datasets for user identification on new devices. This paper focuses on addressing the identification problem with limited data, enabling user identification in diverse scenarios using a dataset containing only a few postures when encountering new devices. To effectively utilize the large-scale dataset from existing devices, we devise two parallel branches to conduct user identification separately on each device. The feature distribution in the auxiliary branch is expected to be more distinct than that in the target branch due to its substantial training data from the auxiliary device. Consequently, we introduce contrastive learning in the feature space to bring different data features of the same user closer together, thereby enhancing identity recognition performance with limited data in the target branch. Specifically, our contributions are as follows:

\begin{itemize}
    \item We proposed an algorithm for user identification with limited data based on contrastive learning. The method allows us to achieve user identification in rich scenarios using a pressure dataset containing only a few postures.
    \item We verified the effectiveness of the algorithm on two 8-subject pressure datasets, BedPressure and ChrPressure. When only using data containing 2 postures for training, our method achieves an average user recognition accuracy of 79.05\% in 12 posture scenes, which is 2.62\% higher than the best baseline model.
\end{itemize}

\section{Related Work}
\label{II}

\subsection{Transfer Learning}
Transfer learning has emerged as a powerful paradigm for addressing the challenges associated with limited labeled data in target domains~\cite{tl1, tl6}. The core idea is to leverage knowledge from related source domains, where abundant labeled data is available, to improve the performance of learning models in the target domains. 

One popular type of transfer learning is domain adaptation, which is the focus of our investigation. Existing domain adaptation methods are roughly divided into three categories: \textbf{(1)} Instance-based methods. Some works~\cite{tl2, tl3, tl4} balance the differences in marginal distributions by assigning weights to source domain instances in the loss function; \textbf{(2)} Feature-based methods. This type of method usually uses autoencoders~\cite{tl5}, adversarial networks, etc. to reconstruct target domain samples to learn Domain-Invariant Features ~\cite{tl7, tl8, tl9}; \textbf{(3)} Parameter-based methods. For example, by pre-training the model on other tasks or other domains and then fine-tuning on the target task or target domain ~\cite{tl10, tl11}. These methods have been widely used in fields such as computer vision and natural language processing.

\subsection{Few-Shot Learning}

Few-shot learning~(FSL)~\cite{fsl1} refers to the challenge of learning latent patterns in data from a limited number of training samples. Meta-learning has emerged as a potent strategy in the research trajectory of FSL, involving training models across multiple related tasks to enable rapid adaptation to new tasks. Metric-based approaches like Matching Networks~\cite{fsl2} and Prototypical Networks~\cite{fsl3} classify by learning sample similarities and computing distances between samples and class centers, demonstrating outstanding performance in few-shot learning. Optimization techniques such as Model-Agnostic Meta-Learning~\cite{fsl4} enables rapid adaptation to task-specific parameters for learning on new tasks. This method focuses on finding the best parameter initialization to speed up model learning. Furthermore, model-based meta-learning approaches such as Meta-Transfer Learning~\cite{fsl5} concentrates on designing specific model architectures tailored for FSL tasks.


Therefore, while FSL also operates on limited data samples, its objective is not centered around training the model to recognize individual samples. Instead, FSL emphasizes learning to differentiate similarities between two samples.

\subsection{Contrastive Learning}

Contrastive learning is a self-supervised learning method that uses triplet loss~\cite{new1}, NCE~\cite{cl4}, or other modified contrastive loss~\cite{cl5, cl9} to project the extracted features into the latent space, where positive pairs are close to each other and negative pairs are far apart. 

Contrastive learning is widely utilized across various disciplines~\cite{cl2, cl3}. Shohreh Deldari~\cite{cl6} and Yuan et al.~\cite{cl7} use contrastive learning to capture similarities between different modalities. Introducing a novel positive-negative pair sampling strategy for multimodal temporal data, they designated data from different modalities at the same timestamp as positive pairs and data from the same modality at different timestamps as negative pairs, and then introduced a new objective function to implement the feature learning task. But this can easily cause class collision problems. Chen et al~\cite{cl8} added a multi-layer perceptron~(MLP) before using contrastive learning, and then used an encoder that dropped the MLP to fine-tune on specific tasks, which significantly improved the performance of contrastive learning. Their experiments show that contrastive learning does not necessarily require operating in a high-dimensional feature space.

Beliz Gunel~\cite{cl10} and Prannay Khosla~\cite{cl11} each extended and refined contrastive loss under supervised learning. They innovatively expanded the notion of positive and negative samples beyond individual instances to encompass entire sample groups based on labels. Inspired by these advancements, we integrated contrastive loss into our proposed method to guide the target domain with blurred boundaries towards the clearly demarcated auxiliary domain.

\section{DataSet}
\label{III}

\begin{figure}[tp]
    \centering
    \subfigure[Smart bed]{
        \includegraphics[width=0.27\textwidth]{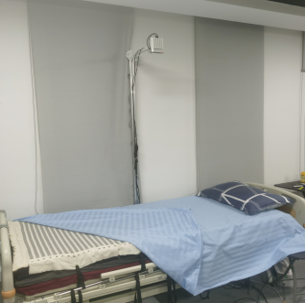}
        \label{fig:setup1}
    }
    \subfigure[Smart chair]{
        \includegraphics[width=0.18\textwidth]{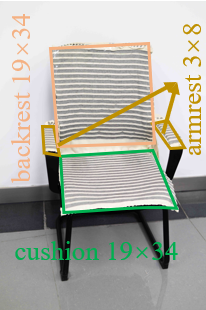}
        \label{fig:setup2}
    }
    \subfigure[RGB images and corresponding pressure images]{
        \includegraphics[width=0.98\linewidth]{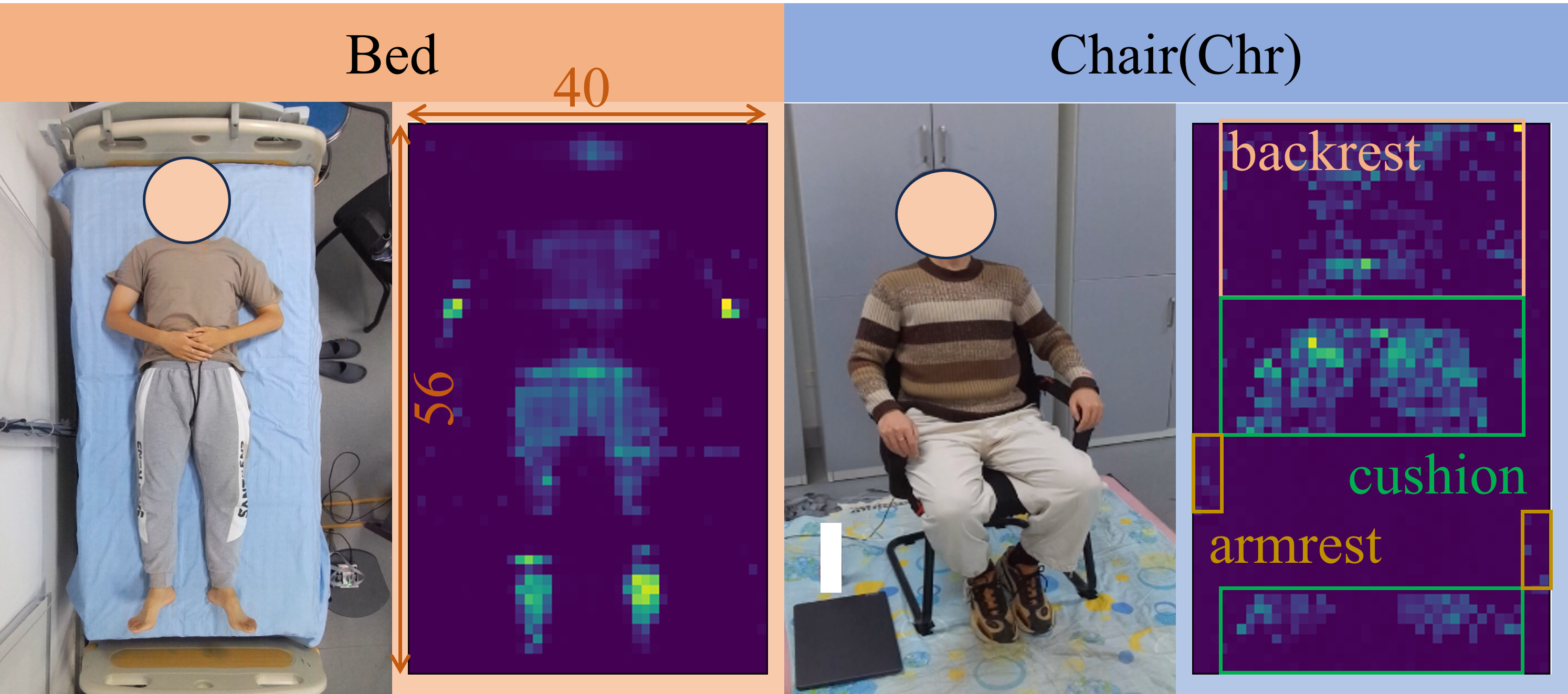}
    }
    \caption{Experimental setup and pressure data information}
    \label{fig:device}
\end{figure}

\begin{figure*}[htbp]
    \centering
    \includegraphics[width=\linewidth]{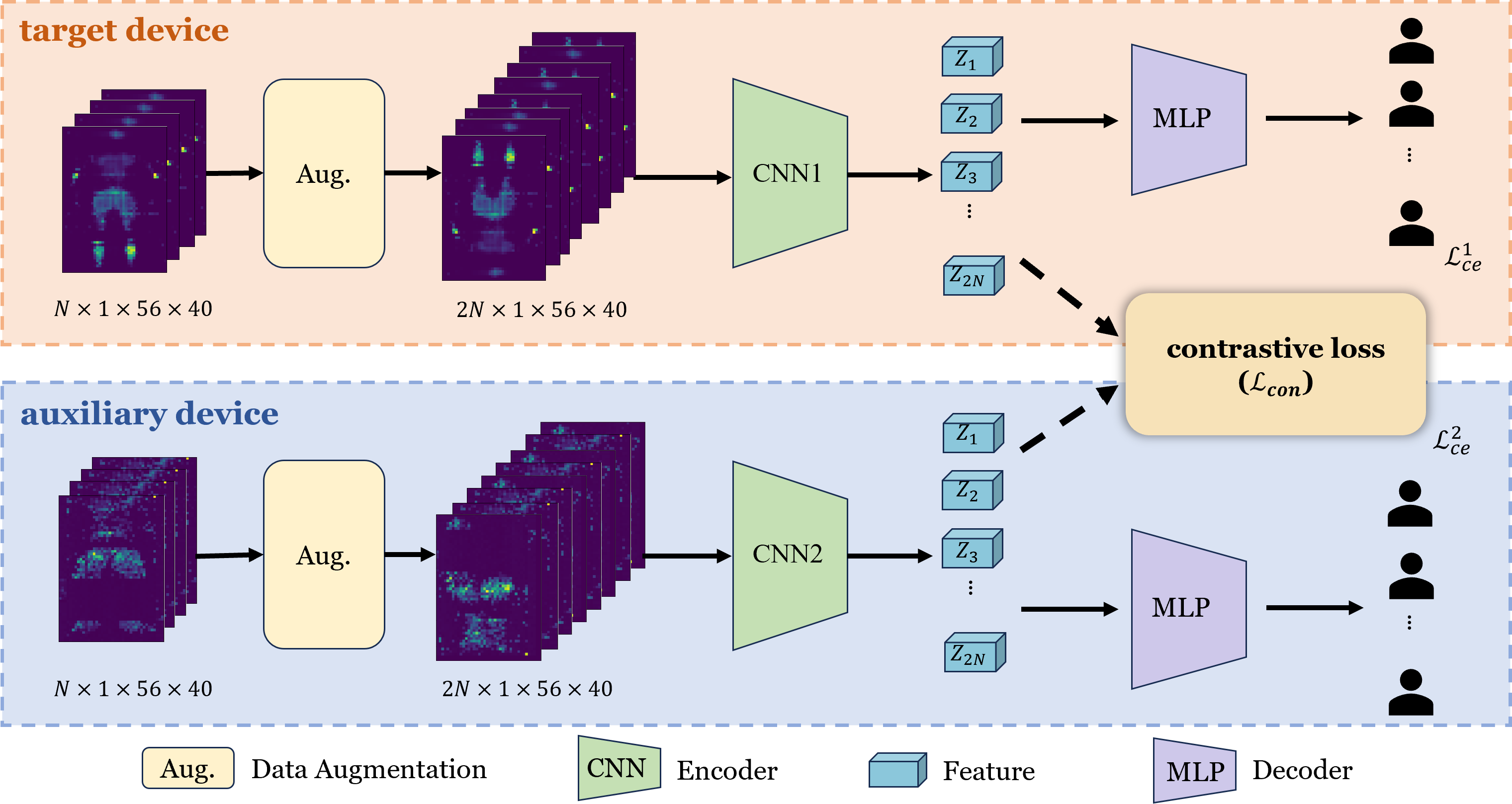}
    \caption{Contrastive learning-based user identification framework}
    \label{fig:framework}
\end{figure*}

Beds and chairs, being the most commonly used devices in our daily life, cover the vast majority of our living scenarios. Therefore, we covered the bed and chair with pressure sensing arrays and used them as collection tools for the datasets used in the experiment~(as shown in Fig.~\ref{fig:setup1} and Fig.~\ref{fig:setup2}). It should be mentioned that our smart bed uses a complete pressure sensing array with a resolution of $56\times40$, while the smart chair is pieced together with a four-part pressure sensing array~(backrest$\times1$, cushion$\times1$, armrest$\times2$) to form $56\times40$. For convenience, we refer to a frame of pressure image as a sample.

Fig.~\ref{fig:device} shows our experimental devices and pressure images collected from them. In our eight-subject data collection experiment, we designed 6 postures on the bed and 12 postures on the chair for data collection. The bed has a frame rate of around 30fps, while the chair has a frame rate of around 20fps. During the data collection process, we specified the overall posture of the subject~(including legs and torso), and allowed the free movement of other parts to ensure the validity of the data frame, and finally obtained two 8-subject pressure datasets, BedPressure and ChrPressure. 

\section{Limited Data-based User Identification}
\label{IV}
\subsection{Overview}\label{AA}

Fig.~\ref{fig:framework} shows the overall framework of our proposed user identification algorithm. For convenience, we refer to new device with limited data for user identification as "target device" and existing device with large dataset as "auxiliary device". In order to fully utilize the large-scale data on auxiliary device, we use an encoder-decoder structure to devise two parallel branches for the target device and the auxiliary device, each of which implements eight-subject identification on the current device. At the same time, we introduce contrastive loss in the feature space to improve the learning ability of the target device when training data is limited. We will provide a detailed overview of the model's design specifics below.

\subsection{Design Details}
\vspace{0.1cm}
\textbf{Encoder-Decoder.} We first denote the input data as $D = \{X_{\text{tar}}, Y_{\text{tar}}, X_{\text{aux}}, Y_{\text{aux}}\}$, where
\begin{align*}
    X_{\text{tar}} & = \{X_{\text{tar}}^{1}, X_{\text{tar}}^{2}, \ldots, X_{\text{tar}}^{N}\} \in \mathbb{R}^{N \times 1 \times 56 \times 40}, \\
    Y_{\text{tar}} & = \{Y_{\text{tar}}^{1}, Y_{\text{tar}}^{2}, \ldots, Y_{\text{tar}}^{N}\} \in \mathbb{R}^N,
\end{align*}
represents the pressure data and labels of the target device~(where $N$ is the batch size). Similarly,
\begin{align*}
    X_{\text{aux}} & = \{X_{\text{aux}}^{1}, X_{\text{aux}}^{2}, \ldots, X_{\text{aux}}^{N}\} \in \mathbb{R}^{N \times 1 \times 56 \times 40}, \\
    Y_{\text{aux}} & = \{Y_{\text{aux}}^{1}, Y_{\text{aux}}^{2}, \ldots, Y_{\text{aux}}^{N}\} \in \mathbb{R}^N,
\end{align*}
represents the pressure data and labels of the auxiliary device~(where $N$ is the batch size).

Then, we define the target encoder as $F_{tar}(\cdot)$, and the decoder as $G_{tar}(\cdot)$. Similarly, we define the auxiliary encoder as $F_{aux}(\cdot)$, and the decoder as $G_{aux}(\cdot)$. At the same time, in order to improve the model's generalization ability and balance the number of positive and negative samples in subsequent contrastive learning, we do not directly use the original input, but introduce a data augmentation module $Aug(\cdot)$. We probabilistically use translation, rotation, and flip operations to generate an enhanced version of the data, which is then fed into the encoder along with the original input. Therefore, we can get the encoded features of the data from the two devices:
\begin{align}
    Z_{tar} = F_{tar}(\{X_{tar}, Aug(X_{tar})\}) \\
    Z_{aux} = F_{aux}(\{X_{aux}, Aug(X_{aux})\})
\end{align}

\begin{table*}[tp]
\caption{2P50S results on ChrPressure Dataset and BedPressure Dataset}
\begin{center}
    \renewcommand{\arraystretch}{1.5}
    \begin{tabular}{cc|ccccccc}
         \Xhline{1.2px}
         \multirow{2}{*}{Auxiliary Device} & \multirow{2}{*}{Methods} & \multicolumn{6}{c}{\textbf{Accuracy on ChrPressure}} & \multicolumn{1}{c}{\textbf{Accuracy on BedPressure}}\\
         & & 1 & 2 & 3 & 4 & 5 & avg & avg\\
         \Xhline{1px}
         $\times$ & KNN & 0.7642 & 0.7300 & 0.6533 & 0.6660 & 0.6856 & 0.6998±0.0644 & 0.7721±0.0322\\
         \Xhline{0.8px}
         $\times$ & ResNet-34+Nil & 0.7485 & 0.7210 & 0.6763 & 0.7204 & 0.7488 & 0.7230±0.0467 & 0.8206±0.0678\\
         $\times$ & ResNet-34+Aug. & 0.7862 & 0.7620 & 0.7484 & 0.7466 & 0.7785 & 0.7643±0.0219 & 0.8438±0.0406\\
         $\times$ & ResNet-34+Recon. & 0.6210 & 0.6679 & 0.5627 & 0.6520 & 0.7191 & 0.6445±0.0818 & 0.8102±0.0772\\
         \Xhline{0.8px}
         $\checkmark$ & ResNet-34+Trans. & 0.6901 & 0.7290 & 0.7372 & 0.7833 & 0.7862 & 0.7452±0.0551 & 0.8233±0.0413\\
         $\checkmark$ & ResNet-34+GAN & 0.7917 & 0.7663 & 0.7354 & 0.7829 & 0.6769 & 0.7506±0.0737 & 0.7204±0.1095 \\
         $\checkmark$ & Ours & \textbf{0.8010} & \textbf{0.7942} & \textbf{0.7538} & \textbf{0.7960} & \textbf{0.8073} & \textbf{0.7905±0.0367} & \textbf{0.8796±0.0378} \\
         \Xhline{1.2px}
    \end{tabular}
    \label{tab: chr}
\end{center}
\end{table*}

In order to allow two parallel branches to complete identification on their respective device, we use the following two independent cross-entropy losses as optimization functions:
\begin{align}
    \mathcal{L}_{ce}^1=CrossEntropy\left(\{Y_{tar},Y_{tar}\},G_{tar}(Z_{tar})\right) \label{eq:lce1}\\
    \mathcal{L}_{ce}^2=CrossEntropy\left(\{Y_{aux},Y_{aux}\},G_{aux}(Z_{aux})\right) \label{eq:lce2}
\end{align}

where $CrossEntropy(\cdot)$ is used to calculate the cross-entropy loss.

\vspace{0.2cm}
\textbf{Supervised Contrastive Learning.} When only independent branches are used, the algorithm becomes a conventional user identification algorithm with data augmentation, lacking sufficient learning capacity especially in scenarios with limited data. Therefore, we use a supervised contrastive loss in the feature space,
\begin{align}
    \mathcal{L}_{con} = \frac1{|I|}\sum_{i\in l}-log\left\{\frac1{|P(i)|}\sum_{p\in P(i)}\frac{\exp(Z_i\cdot Z_p)/\tau)}{\sum_{a\in A(i)}\exp(Z_i\cdot Z_a/\tau)}\right\}
    \label{eq:lcon}
\end{align}

where $I$ represents the sample set corresponding to the feature set $\{Z_{tar}, Z_{aux} \}$, $P(i)$ represents the positive sample set of sample $i$, $A(i)$ represents the sample set without $i$, and $\tau$ is a temperature parameter.

This contrastive loss means that we draw on the high classification accuracy of large-scale data on auxiliary device to move positive samples closer to each other and negative samples farther from each other in the feature space. Relying on the contrastive loss, the target device's encoder can better infer other postures that users might assume~(even when our training dataset has limited posture diversity). This represents an intuitive endeavor to bolster the model's generalization capability.

\vspace{0.2cm}
\textbf{Loss Function.} Finally, we use the weighted sum of the three parts of the loss~(~\ref{eq:lce1}, ~\ref{eq:lce2}, ~\ref{eq:lcon}) as the loss function of the model,
\begin{align}
    \mathcal{L} = \lambda_{1}\mathcal{L}_{ce}^{1}+\lambda_{2}\mathcal{L}_{ce}^{2}+\lambda_{3}\mathcal{L}_{con}
    \label{eq:loss}
\end{align}

where $\lambda$ is the weight of each loss.

\section{Experiment}
\label{V}

In order to describe the limited data used for training, we introduce the concept of "mPnS", where "mP" means that the dataset used for training only contain m postures; "nS" means that the data in each posture only contain n data samples. In order to verify the comprehensive performance of our proposed method, we conducted the following experiments.

\subsection{User Identification on ChrPressure Dataset}

To verify the identification accuracy of the algorithm for limited data, we initially employ grid search to establish the following parameters: $\tau = 0.10$, $\lambda_1 = 0.15$, $\lambda_2 = 0.15$, and $\lambda_3 = 0.7$. We utilize ResNet-34 as the encoder, set the batch size to 32, and fix the learning rate at 0.0005 for training over 150 epochs, and then compare with the following six baseline models in Table~\ref{tab:baseline} under the experimental settings of 2P50S.

\begin{table}[htbp]
    \centering
    \caption{Baseline Models for Comparison}
    \renewcommand{\arraystretch}{1.7}
    \label{tab:baseline}
    \begin{tabular}{m{2.2cm}|m{5.5cm}}
        \Xhline{1.2px}
        \textbf{Baseline Models} & \textbf{Description}\\
        \Xhline{1.0px}
        KNN & K-Nearest Neighbor\\
        \Xhline{0.8px}
        ResNet-34+Nil & A pure model without using any strategy.\\
        \Xhline{0.8px}
        ResNet-34+Aug. & A pure model with added data augmentation.\\
        \Xhline{0.8px}
        ResNet-34+Recon. & Pretrain the encoder using an image reconstruction task on the target dataset.\\ 
        \Xhline{0.8px}
        ResNet-34+Trans. & Transfer learning methods that are pre-trained on large-scale auxiliary data and then fine-tuned on the target dataset.\\
        \Xhline{0.8px}
        ResNet-34+GAN & Add a discriminator to the dual-branch structure and use adversarial training methods to promote the unification of feature domains.\\
        \Xhline{1.2px}
    \end{tabular}
\end{table}

To eliminate randomness, we conducted five experiments, each of which randomly divided the training set, validation set, and test set. Table~\ref{tab: chr} shows the results of our five experiments. Our method can achieve the best average performance of 79.05\% under the experimental setting of 2P50S, and shows strong stability. In addition, it is noted that the pre-training method based on reconstruction has poor effect. This may be because the chair has multiple sensing areas. The final pressure image is pieced together from these sensing areas, and the semantic information they preserve is extremely weak, which is also confirmed in Experiment~\ref{B}.

\begin{figure}[htbp]
    \centering
    \includegraphics[width=\linewidth]{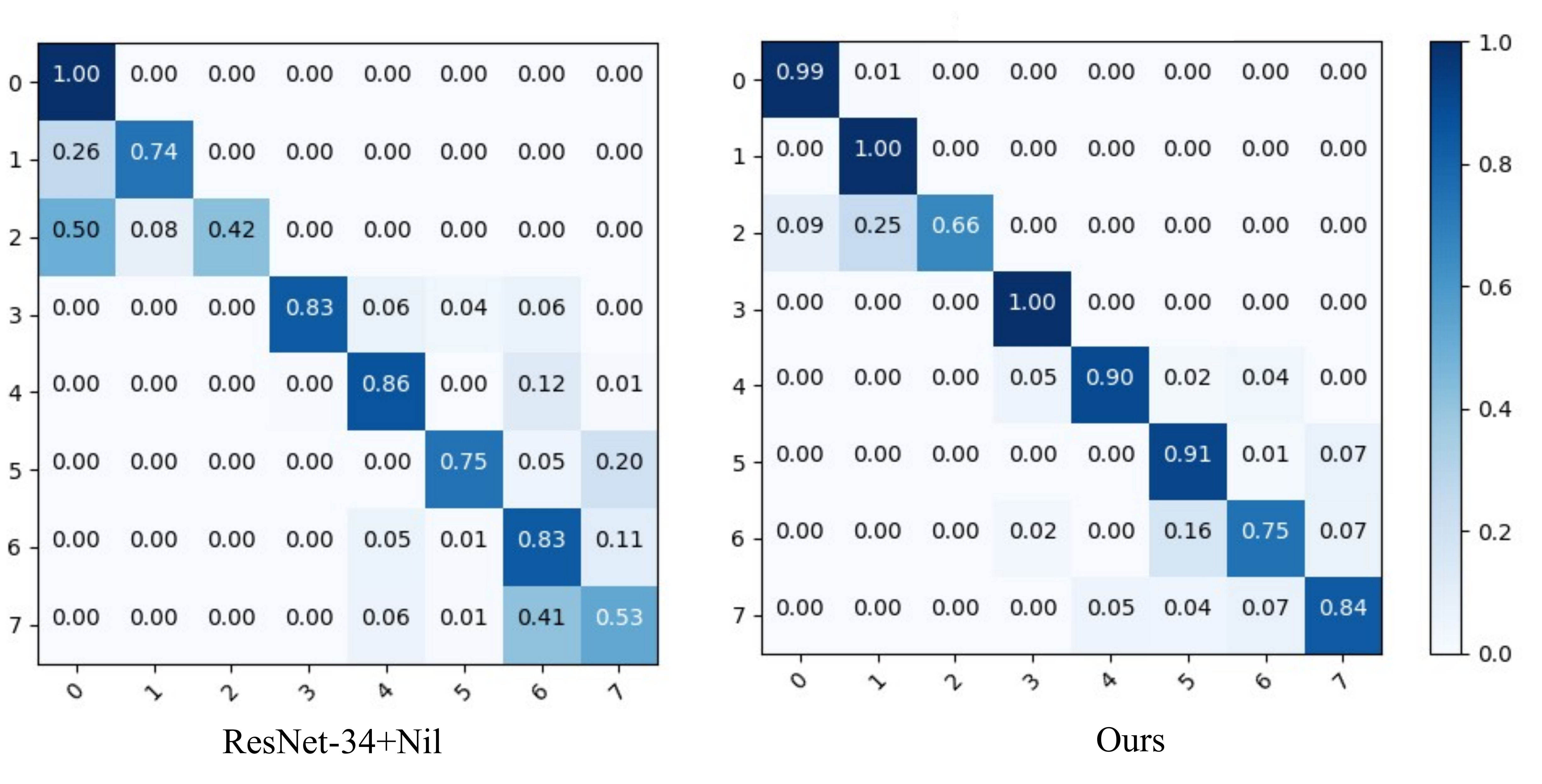}
    \caption{Confusion matrix for ResNet-34+Nil~(left) and Ours~(right)}
    \label{fig:con}
\end{figure}

ResNet-34+Nil demonstrates a typical error that is common to all methods. We will now illustrate this using the confusion matrix in Fig.~\ref{fig:con}. It is evident that ResNet-34+Nil has a lot of confusion between Subject 1, 2, 3 and Subject 6, 7; from the practical results, most of these errors occur between the same postures of different users, which confirms what we mentioned above; although our method also has a small number of these errors, it overall greatly improves the performance of the model in this aspect, which is the key to achieving user identification on low-richness dataset. 

\subsection{User Identification on BedPressure Dataset}
\label{B}
To validate the algorithm's stability in performance relative to other methods on different dataset, we swapped the target and auxiliary device and conducted tests using the 2P50S experimental setup on the BedPressure dataset. Table~\ref{tab: chr} also presents the average results of five random experiments on BedPressure.

From the table~\ref{tab: chr}, it can be found that our results can still maintain the best performance, which shows that our method is expected to be extended to other dataset and has potential scalability; in addition, it is obvious that the results on BedPressure are generally better than ChrPressure. This is because there are only 6 postures per person on BedPressure. Under the same 2P50S settings, it is easier for the model to learn the identity of unknown postures. 

Unlike on ChrPressure, the reconstruction-based method achieves a performance similar to ResNet-34+Nil. The result exceeds the performance on ChrPressure. This is because the pressure image on bedsheet is a complete sensing area, and the reconstruction method is expected to learn the semantic information in it.

\subsection{Ablation Study}\label{SCM}

\subsubsection{The Number of Postures}

\begin{figure}[htbp]
    \centering
    \includegraphics[width=\linewidth]{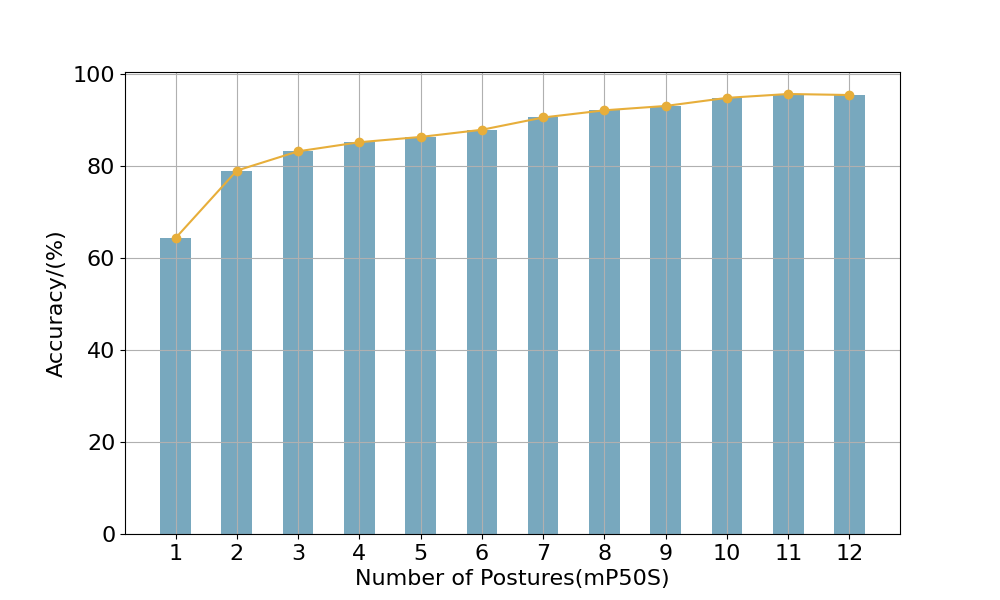}
    \caption{User identification accuracy under different number of postures}
    \label{fig:ab1}
\end{figure}

The number of postures~(i.e., m in mPnS) is an important criterion for measuring the tediousness of data collection work. Fig.~\ref{fig:ab1} shows the accuracy of our method on ChrPressure as a function of the number of postures used for training.

It can be found that the accuracy of our method increases with the addition of posture categories in Fig.~\ref{fig:ab1}. Obviously, this outcome aligns with expectations, as the growth in the number of postures implies higher dataset richness, which is conducive to the learning of the feature. However, this simultaneously introduces challenges in data collection, particularly for smart textile device that prioritize user experience. Therefore, determining the number of actions should strike a balance between accuracy and user experience.

\subsubsection{The Sample Number of Each Posture}

\begin{figure}[htbp]
    \centering
    \includegraphics[width=\linewidth]{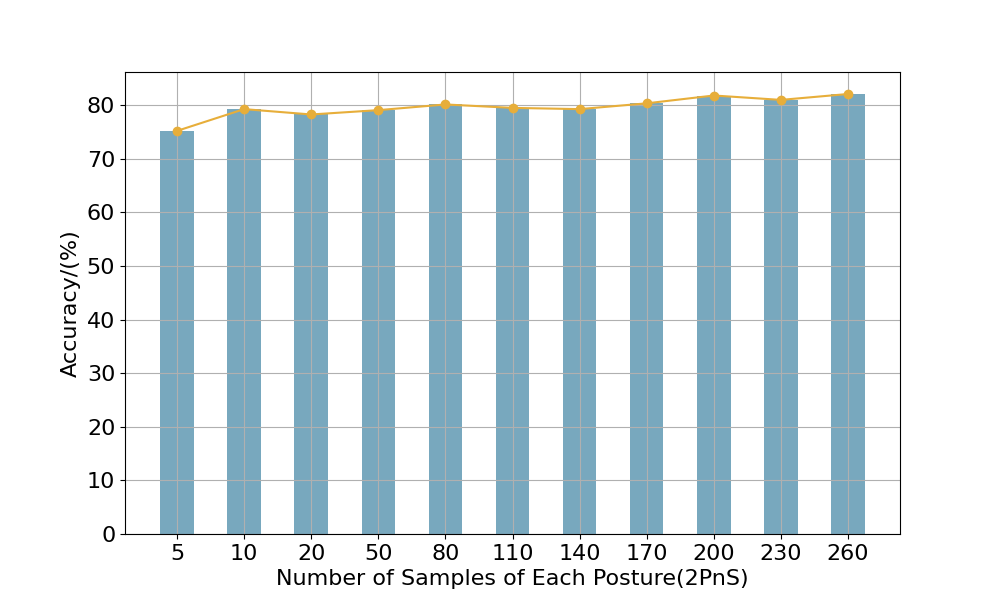}
    \caption{User identification accuracy under different size of each posture}
    \label{fig:ab2}
\end{figure}

The sample number of each posture~(i.e. n in mPnS) is an important measure of how long the data collection effort takes. Fig.~\ref{fig:ab2} shows the accuracy of our method on ChrPressure as a function of the size of each posture used for training.

It can be found that, although we allow some subtle movements during the data collection process to ensure detailed changes in pressure frames, increasing the number of samples used in each posture does not lead to a significant change in the results. In fact, based on the sampling frequency of the current device, it only takes about 2 seconds to collect 50 frames of data, so fewer data samples do not have stronger practical significance. 

\subsubsection{Encoder-decoder implementation}

\begin{table}[bp]
    \centering
    \caption{User identification accuracy on chrpressure under different implementation of Encoder-Decoder}
    \renewcommand{\arraystretch}{1.7}
    \label{tab:enc-dec}
    \begin{tabular}{c|c|c}
        \Xhline{1.2px}
        Encoder & Decoder & Accuracy \\
        \Xhline{1.0px}
        \multirow{2}{*}{ResNet-18} & Shared & 0.7433\\
         & Independent & 0.7789\\
        \Xhline{0.8px}
        \multirow{2}{*}{ResNet-34} & Shared & 0.7856\\
         & Independent & 0.8073\\
        \Xhline{0.8px}
        \multirow{2}{*}{ResNet-50} & Shared & 0.7992\\
         & Independent & 0.8183\\
        \Xhline{1.2px}
    \end{tabular}
\end{table}

In practice, the choice of encoder is not fixed and depends on the trade-off between accuracy and computational cost. In addition, parallel designed decoders also affect the performance of the model. To explore the impact of different designs on the results, we conducted ablation studies on different encoder implementations and independent/common decoder designs.

Table~\ref{tab:enc-dec} shows our experimental results. As we just discussed, it turns out that independently designed decoders tend to have better results than shared decoder. Changing the structure of the encoder may bring about changes in results. The specific choice should depend on the balance between computational cost and model performance.

\section{CONCLUSIONS}

\label{VI}

In this paper, we present a user identification method specifically tailored for limited data using contrastive learning. This approach is intended for use in multi-device environments with pressure-sensitive smart devices. When encountering a new device, our method only requires data collection from a few basic postures to enable user identification across at least 12 postures. Furthermore, the abundant data available on existing devices can be utilized to support user identification efforts. Experimental results demonstrate that under 2P50S settings, the proposed method achieves an average performance of 79.05\% on ChrPressure dataset and 87.96\% on BedPressure dataset, outperforming all other models. We also explore the impact of different numbers of postures and samples on the results and provide recommendations for balancing data acquisition cost and accuracy.

\addtolength{\textheight}{-4cm}   










\begin{thebibliography}{99}
\bibitem{id3} Zhou Z, Padgett S, Cai Z, et al. Single-layered ultra-soft washable smart textiles for all-around ballistocardiograph, respiration, and posture monitoring during sleep[J]. Biosensors and Bioelectronics, 2020, 155: 112064.
\bibitem{b1} Zhou B, Suh S, Rey V F, et al. Quali-mat: Evaluating the quality of execution in body-weight exercises with a pressure sensitive sports mat[J]. Proceedings of the ACM on Interactive, Mobile, Wearable and Ubiquitous Technologies, 2022, 6(2): 1-45.
\bibitem{luo1} Luo Y, Wu K, Palacios T, et al. KnitUI: Fabricating interactive and sensing textiles with machine knitting[C]//Proceedings of the 2021 CHI Conference on Human Factors in Computing Systems. 2021: 1-12.
\bibitem{id6} Davoodnia V, Slinowsky M, Etemad A. Deep multitask learning for pervasive bmi estimation and identity recognition in smart beds[J]. Journal of Ambient Intelligence and Humanized Computing, 2023, 14(5): 5463-5477.

\bibitem{luo2} Luo Y, Li Y, Foshey M, et al. Intelligent carpet: Inferring 3d human pose from tactile signals[C]//Proceedings of the IEEE/CVF conference on computer vision and pattern recognition. 2021: 11255-11265.
\bibitem{b6} Nawawi M M M, Sidek K A, Dafhalla A K Y, et al. Review on data acquisition of electrocardiogram biometric recognition in wearable smart textile shirts[C]//Journal of Physics: Conference Series. IOP Publishing, 2021, 1900(1): 012019.
\bibitem{b7} Zhang X, Ye H, Peng T, et al. A Smart User Authentication Approach using Sensing Seat[C]//2020 IEEE 16th International Conference on Automation Science and Engineering (CASE). IEEE, 2020: 317-322.
\bibitem{tl1} Zhuang F, Qi Z, Duan K, et al. A comprehensive survey on transfer learning[J]. Proceedings of the IEEE, 2020, 109(1): 43-76.

\bibitem{tl6} Farahani A, Voghoei S, Rasheed K, et al. A brief review of domain adaptation[J]. Advances in data science and information engineering: proceedings from ICDATA 2020 and IKE 2020, 2021: 877-894.

\bibitem{tl2} Huang J, Gretton A, Borgwardt K, et al. Correcting sample selection bias by unlabeled data[J]. Advances in neural information processing systems, 2006, 19.
\bibitem{tl3} Sun Q, Chattopadhyay R, Panchanathan S, et al. A two-stage weighting framework for multi-source domain adaptation[J]. Advances in neural information processing systems, 2011, 24.
\bibitem{tl4} Jiang J, Zhai C X. Instance weighting for domain adaptation in NLP[C]. ACL, 2007.
\bibitem{tl5} Zhao M, Yue S, Katabi D, et al. Learning sleep stages from radio signals: A conditional adversarial architecture[C]//International Conference on Machine Learning. PMLR, 2017: 4100-4109.
\bibitem{tl7} Yang S, Yu K, Cao F, et al. Dual-representation-based autoencoder for domain adaptation[J]. IEEE Transactions on Cybernetics, 2021, 52(8): 7464-7477.

\bibitem{tl8} Wang X, Ma Y, Cheng Y, et al. Heterogeneous domain adaptation network based on autoencoder[J]. Journal of Parallel and Distributed Computing, 2018, 117: 281-291.

\bibitem{tl9} Choi J, Kim T, Kim C. Self-ensembling with gan-based data augmentation for domain adaptation in semantic segmentation[C]//Proceedings of the IEEE/CVF international conference on computer vision. 2019: 6830-6840.
\bibitem{tl10} Li J, He R, Ye H, et al. Unsupervised domain adaptation of a pretrained cross-lingual language model[J]. arXiv preprint arXiv:2011.11499, 2020.
\bibitem{tl11} Diao S, Xu R, Su H, et al. Taming pre-trained language models with n-gram representations for low-resource domain adaptation[C]//Proceedings of the 59th Annual Meeting of the Association for Computational Linguistics and the 11th International Joint Conference on Natural Language Processing (Volume 1: Long Papers). 2021: 3336-3349.
\bibitem{fsl1} Wang Y, Yao Q, Kwok J T, et al. Generalizing from a few examples: A survey on few-shot learning[J]. ACM computing surveys (csur), 2020, 53(3): 1-34.

\bibitem{fsl2} Vinyals O, Blundell C, Lillicrap T, et al. Matching networks for one shot learning[J]. Advances in neural information processing systems, 2016, 29.
\bibitem{fsl3} Snell J, Swersky K, Zemel R. Prototypical networks for few-shot learning[J]. Advances in neural information processing systems, 2017, 30.
\bibitem{fsl4} Finn C, Abbeel P, Levine S. Model-agnostic meta-learning for fast adaptation of deep networks[C]//International conference on machine learning. PMLR, 2017: 1126-1135.
\bibitem{fsl5} Sun Q, Liu Y, Chua T S, et al. Meta-transfer learning for few-shot learning[C]//Proceedings of the IEEE/CVF conference on computer vision and pattern recognition. 2019: 403-412.
\bibitem{new1} Weinberger K Q, Saul L K. Distance metric learning for large margin nearest neighbor classification[J]. Journal of machine learning research, 2009, 10(2).
\bibitem{cl4} Gutmann M, Hyvärinen A. Noise-contrastive estimation: A new estimation principle for unnormalized statistical models[C]//Proceedings of the thirteenth international conference on artificial intelligence and statistics. JMLR Workshop and Conference Proceedings, 2010: 297-304.
\bibitem{cl5} Tian Y, Krishnan D, Isola P. Contrastive multiview coding[C]//Computer Vision–ECCV 2020: 16th European Conference, Glasgow, UK, August 23–28, 2020, Proceedings, Part XI 16. Springer International Publishing, 2020: 776-794.
\bibitem{cl9} Zheng M, Wang F, You S, et al. Weakly supervised contrastive learning[C]//Proceedings of the IEEE/CVF International Conference on Computer Vision. 2021: 10042-10051.
\bibitem{cl2} Shen X, Liu X, Hu X, et al. Contrastive learning of subject-invariant eeg representations for cross-subject emotion recognition[J]. IEEE Transactions on Affective Computing, 2022.
\bibitem{cl3} Bhati S, Villalba J, Żelasko P, et al. Unsupervised speech segmentation and variable rate representation learning using segmental contrastive predictive coding[J]. IEEE/ACM Transactions on Audio, Speech, and Language Processing, 2022, 30: 2002-2014.
\bibitem{cl6} Deldari S, Xue H, Saeed A, et al. Cocoa: Cross modality contrastive learning for sensor data[J]. Proceedings of the ACM on Interactive, Mobile, Wearable and Ubiquitous Technologies, 2022, 6(3): 1-28.
\bibitem{cl7} Yuan X, Lin Z, Kuen J, et al. Multimodal contrastive training for visual representation learning[C]//Proceedings of the IEEE/CVF Conference on Computer Vision and Pattern Recognition. 2021: 6995-7004.
\bibitem{cl8} Chen T, Kornblith S, Norouzi M, et al. A simple framework for contrastive learning of visual representations[C]//International conference on machine learning. PMLR, 2020: 1597-1607.
\bibitem{cl10} Gunel B, Du J, Conneau A, et al. Supervised contrastive learning for pre-trained language model fine-tuning[J]. arXiv preprint arXiv:2011.01403, 2020.
\bibitem{cl11} Khosla P, Teterwak P, Wang C, et al. Supervised contrastive learning[J]. Advances in neural information processing systems, 2020, 33: 18661-18673.


\end{thebibliography}
\end{document}